\documentclass[jkps,preprint,fleqn,showpacs,showkeys]{revtex4}
\usepackage{graphicx}
\usepackage{epstopdf}
\usepackage{amssymb}
\usepackage{amsmath}
\usepackage{bm}
\usepackage[font=footnotesize]{subfig}
\begin{document}
\setcounter{page}{0}
\title[]{Optimization of the Multi-turn Injection Efficiency for Medical Synchrotron}
\author{J. \surname{Kim}}
\affiliation{Department of Physics,  Pohang University of Science and Technology,
77 Cheongam-Ro, Nam-Gu, Pohang, Gyeongbuk, Korea 790-784 }
\author{H. \surname{Yim}}
\affiliation{KIRAMS, 75 Nowon-ro, Nowon-gu, Seoul, Korea  139-706}
\author{M. \surname{Yoon}}
\affiliation{Department of Physics,  Pohang University of Science and Technology,
77 Cheongam-Ro, Nam-Gu, Pohang, Gyeongbuk, Korea 790-784 }
\email{moohyun@postech.ac.kr}
\thanks{Fax: +82-54-279-3099 }

\date[]{}

\begin{abstract}
We present a method for optimization of the
multi-turn injection effciency for medical synchrotron. We show
that for given injection energy the injection efficiency can be
greatly enhanced by choosing transverse tunes appropriately
as well as optimizing the injection bump and the number of turns
required for beam injection.
We verify our study by applying
the method to the Korea Heavy Ion Medical Accelerator (KHIMA)
synchrotron which is currently built at
the campus of Dongnam Institute of Radiological and Medical Sciences
(DIRAMS) in Busan, Korea. First the frequency map analysis is
performed with the help of ELEGANT and ACCSIM codes. The tunes
which yield the good injection efficiency are then selected. With
these tunes the injection bump and the number of turns required for injection
are then optimized by tracking a number of particles up to one
thousand turns after injection beyond which there is no further beam loss. Results for
optimization of the injection efficiency for proton ion are presented.

\end{abstract}



\maketitle

\section{INTRODUCTION}
One of the promising methods for treatment of malignant tumors is the particle beam therapy.
Specifically, there are numerous positive reports on effectiveness and prognosis of ion beam therapy
as a non surgical treatment~\cite{Okada}
using proton or heavier ion such as a carbon ion, and as a result the ion beam therapy is
increasingly popular worldwide.

After Bragg peak property of an ion beam was found to be
better than the conventional photon beam in curing patients with tumor,
first clinical experience on a human being with a proton beam was reported by researchers
in University of California, Berkeley in 1954~\cite{Giap}.
Soon after that many proton facilities were developed around the world.
In 1990, the world first hospital-based proton therapy facility was constructed in Loma Linda, USA~\cite{Giap}.
So far there are about forty operating proton facilities worldwide and
the total number of patients treated by particle beam therapy exceeded 100,000~\cite{Giap}.
Success of these facilities stimulated the construct of more ion beam facilities.

In 1996, design study of a proton-ion beam accelerator for tumor treatment
began to explore the possibility of building such a facility in Europe~\cite{Pimms}.
The aim of this study, called Proton-Ion Medical Machine Study (PIMMS), is to design
a compact medical accelerator for acceleration of a proton beam and a carbon beam. The result was
reported in 1999~\cite{Pimms}.

Of particular importance of this design is a systematic study of beam injection into
and extraction from a synchrotron. An issue associated with the injection of an ion beam is
the space-charge effect which is especially pronounced at low energy
of a beam. It is a repulsive force due to Coulomb interaction of charged particles that  is
inversely proportional to $\gamma^2$ where $\gamma$ is the usual relativistic factor.
This means that high energy of injected beam is preferred, but high injection energy
means high cost of the injector accelerator, usually linear accelerator. Therefore,
an optimization study of the injection energy is necessary. Such study has been performed
extensively during PIMMS~\cite{Pimms}; the injection energy in PIMMS is 7 MeV/u.

Another important aspect related with the beam injection into a synchrotron
is to maximize the injection efficiency. In any event maximization of the injection efficiency
has to be achieved in particle accelerators. However,
in a medical synchrotron employing multi-turn injection this is particularly important
because the beam is accumulated turn by turn until desired beam current is reached. In the multi-turn
injection scheme, injection is achieved with the help of time-varying bump magnets.
The phase space occupied by injected beams in the injection plane (usually horizontal plane)
is diluted after injection (see for example Figs. 3, 6, 7, 8) and therefore the effective beam
emittance in the injection plane
is very large, usually around 100 $\pi$ mm mrad.  Although the acceptance of a synchrotron is
designed in such a way that it is greater than the effective beam emittance, beam loss
can occur immediately after injection or during acceleration due to various reason. Injection
and extraction septa cause the main limitation in defining the acceptance of synchrotron.
It is of utmost importance to design an injection scheme carefully to minimize the beam loss.

The purpose of the present paper is to investigate a way to enhance the beam injection efficiency
in a synchrotron employing the multi-turn injection. Especially we focus on searching optimal
transverse tunes at injection which maximize the injection efficiency. Assuming linear
rise and fall of the bump magnets' pulse, effect of the pulse duration and bump angle on the injection is
also examined.  To verify our study we take the synchrotron
lattice from Korea Heavy Ion Medical Accelerator (KHIMA) currently built in Busan,
which is the first medical accelerator
for heavy ion therapy in Korea. In many aspects, KHIMA
follows the PIMMS design.  It has a capability of accelerating proton ion or carbon ion;
different injection linacs for each ion. More details of the KHIMA synchrotron is
described elsewhere~\cite{Yim}.

This paper focuses on the proton injection instead of carbon ion. It is mainly because
for the same injection energy per nucleon the space-charge effect is more pronounced for
proton than carbon ions such as $^{12}$C$^{6+}$.

This paper is organized as follows: In Section 2, the multi-turn injection scheme in KHIMA synchrotron
is introduced. Section 3 presents the tune survey so that dependence of transverse tunes on the
injection efficiency is shown. Section 4 describes the effect of bump magnet's angle
and the pulse duration.
Finally, Conclusion is given.

\section{MULTI-TURN INJECTION}
The KHIMA synchrotron's circumference is 75 m and it has two straight sections with zero dispersion.
An electrostatic injection septum (EIS) and a magnetic extraction septum
are located respectively in these two straight sections.
Table I shows baseline parameters for injection.
\begin{table}
\caption{Parameters for Injection}
\begin{ruledtabular}
\begin{tabular}{lc}
Parameters & Value \\
\hline
Nominal energy & 7\ MeV/u \\
$\eta_x/\eta_y$ at the injection point         &  0 /0  \\
$\eta'_x/\eta'_y$ at the injection point         &  0 /0  \\
Horizontal and vertical position  & 48.5 mm / 4.5 mm \\
\quad of the injection point    \\
Horizontal rms emittance  & 1 $\pi$ mm mrad  \\
Vertical rms emittance    & 1 $\pi$ mm mrad \\
\end{tabular}
\end{ruledtabular}
\end{table}

so that they have same values with the values of synchrotron at the injection point.
The multi-turn injection scheme makes it possible to stack  the horizontal machine
acceptance with successive injection of beams from the injector linac~\cite{Ree}.
Applying Liouville's theorem when a number of beams are injected by turn by turn, injected beams
in later turn should be injected to local phase-space region which is not filled with injected
beams in earlier turn. This is feasible when each beam that is injected in different order
follows different closed orbit path because the injection point is fixed. The closed orbit
path is controlled with the help of bump magnets which are capable of moving the closed
orbit path away from the design orbit (Fig.~1) Setting bump magnets with linearly
decreasing kick angle is one of proper ways to stack each injected beam in order.
The beam injected in the first turn goes through the most long, distorted orbit and
rotates by $2\pi \nu_{x}$ radian in the phase space until the second beam is injected.
At the same time, the closed orbit and the first beam are moving toward the design
orbit as the second beam comes in.
After repeated turns for injection, the machine acceptance is occupied densely
and the closed orbit follows the design orbit.
In other words, when the multi-turn injection is finished, composition of the
machine acceptance is that a beam injected earlier occupies inner region
and a beam injected later occupies outer region in sequence.

Many parameters affect how the injection scheme works ; betatron tunes, initial
kick angle of bump magnets, number of turns for injection, angle of injected beam, twiss parameters of
injected beam, and so on.  Finding the best combination of those parameters is difficult because it is
 a multi-dimensional optimization problem.

In this study, we mainly focus on the effect of betatron tunes and parameters associated with
the bumped orbit such as kick angle and pulse duration of bump magnets.
In this paper, the injection efficiency is defined as the percentage of number of survived particles
divided by number of injected particles.
 \begin{equation}
 {\rm Injection\;efficiency} = \frac{N_{\rm survived}}{N_{\rm injected}}\times 100 (\%)
 \end{equation}

In this investigation 1000 macro particles are injected per each turn with rms horizontal and vertical beam
emittances of 1 $\pi$ mm mrad and the total number of turns for injection
is set to be 20 for tune survey. Then the number of turns for injection
(i.e. bump-pulse duration) is varied to examine its effect on the injection efficiency.
Number of survived particles is counted after the 20-turn injection and then additional 1000 turns
are tracked to count still remaining particles before acceleration.
ELEGANT code~\cite{elegant} was utilized for tune matching and ACCSIM code~\cite{accsim}
was then used for particle tracking during and after injection.

\section{TUNE SURVEY}
For later comparison the initial tunes are chosen to be ($\nu_x$ = 1.73 and $\nu_y$ =  1.47)
and various beam dynamics studies have been performed with these tunes~\cite{Yim}.
Also, the initial kick angle of the bump magnets is 4.8 mrad and the kick angle collapses to zero in 20 turns.
With these parameters, the injection efficiency was calculated to be 33.5 $\%$ which is the baseline
efficiency to be compared with.

Search for optimal horizontal and vertical tunes is carried out by varying tunes in steps of 0.01
from $\nu_x, \ \nu_y$ = 1 to 2. Per each tune, zero dispersion in the straight sections is maintained while
chromaticities are kept constant. Also Twiss parameters at the injection point are maintained for each tune.

Space-charge effect is not included during the computation because
if the space-charge is included the computing time increases significantly. However, the effect of the
space charge on the injection efficiency
was examined once the optimal tunes were selected as shown in the below.

Figure 2 shows the footprint for injection efficiencies plotted on the tune diagram with resonance
lines up to third-order. Poor injection efficiencies around the third-order resonance line
$\nu_{x}+2\nu_{y}$ = 3 can be seen.
From this figure, we have selected (1.82, 1.30) for horizontal and
vertical tunes. At these tunes, the injection efficiency is increased to 41.5 $\%$
and therefore the new tunes
yield eight percentage-point increase in the injection efficiency.

Figure 3 shows normalized horizontal phase-space portraits
($p_x = \beta_x x' + \alpha_x x$ where $\alpha_x$ and $\beta_x$ are the
usual twiss parameters) (a) for initial tunes (1.73, 1.47) and (b)
for new tunes (1.82, 1.30). Comparing these plots it is seen clearly that injected
particles with higher injection
efficiency occupy denser and smaller phase-space area, thus verifying the
superiority of the new tunes. Note that not all particles are survived after injection
as Fig. 3 indicates the total number of pulses is less than 20.

\section{OPTIMIZATION OF THE BUMP MAGNET PARAMETERS}
To explore possibility of enhancing the injection efficiency further, dependence of
the initial kick angle and the pulse duration of the bump magnets on the injection efficiency
were investigated with the new tunes. In KHIMA synchrotron there are two injection bump magnets
and one electrostatic injection septum, in addition to one electrostatic extraction septum.
The phase advance between the bump magnets is approximately $\pi$ radian, so the bump is closed.
Horizontal aperture limit of the EIS is 41 mm. However, the limitation
of the aperture in the KHIMA synchrotron is set by electrostatic extraction septum (EES), which
is 35 mm; most of the particle loss occurs at the position of this septum.

If there is only one septum whose horizontal aperture limits the acceptance of an
accelerator,  it is desirable for linear bump (i.e. rise and fall of the bump pulse are
linear) for injection to satisfy the following condition:
\begin{equation}
\delta_{\rm bump} = \nu_{\rm frac} (2a +d_{\rm sep}),
\end{equation}
where $\delta _{\rm bump}$ is the orbit shift per one revolution,
$\nu_{\rm frac}$ for fractional tune in the injection plane,  $a$
the radius of the injected beam at the position of the septum, and $d_{\rm sep}$ is
the effective septum thickness~\cite{Daqa}. This equation implies that
the pulse duration (or slope of the pulse collapse) of the linearly falling
bump magnet affects the injection efficiency. Also the maximum kick angle of the
bump magnet should be made such that the bumped orbit is close to the injected beam
to minimize the betatron oscillation amplitude of the injected beam.

In our study, we have searched the optimal initial
kick angle of the bump magnets and number of turns needed for the bump
to decrease to zero. The result is shown in Fig. 4 where the footprint for
injection efficiencies as functions of initial bump angle and the bump duration.
The new betatron tunes were assumed,
i.e. (1.82, 1.30). Figure 4 indicates that there is a large island region showing higher
injection efficiency. We have chosen 4.2 mrad for the initial angle and 27
for the number of turns to collapse.
In obtaining Fig. 4 the injection is terminated at the 20$^{th}$ turn.
The reason why the injection is terminated earlier than the turn number
at which the bump collapse to zero is depicted in Fig. 5. This figure shows
the number of accumulated particles as a function of the number of turns
for old (plus marks) and new (cross marks) tunes and bump parameters mentioned above.
The shaded bars in this figure indicate the number of desired accumulated macro particles
(i.e. number of particles for 100$\%$ injection efficiency) injected per each turn.
From this figure, one can see that before the improvement total number of turns
required for injection seems to be around 15$^{th}$ because after this turn the
number of accumulated particles does not increase further, meaning that the
rate of the beam loss balances the rate of the injection.

The figure also
shows that the number of turns required for injection after the improvement
(i.e. with new tunes and bump parameters) is around 17.
With the newly found tunes (1.82, 1.30) and the new bump parameters
the injection efficiency was calculated to be 50.5$\%$. This improvement is to be
compared with the old efficiency, 33.5$\%$.
With these new parameters, the normalized horizontal phase space at the
1000$^{th}$ turn after the bump collapse is shown in Fig. 6. When this figure is
compared with Fig. 3(a) and (b), one can see that the new phase portraits
shows better behavior of beam stacking (i.e. beam is centered better).

For proton therapy, the required number of particles to a patient is
1.5$\times$10$^{10}$ per a spill.  We assigned a total of 6$\times$10$^6$ real
particles for a single macroparticle and 1000 macroparticles are used in
the simulation study. With 20 turns for injection together with taking into
consideration of losses during capture, acceleration and extraction,
this number is sufficient even with the old tunes:
for accumulated particles after the injection, we expect 10\% additional loss
during the capture, 20\% additional loss during the acceleration and 10\% additional
loss during the extraction. Hence, we predict the number of survival real
particles per a spill as 6$\times$10$^6$ particles/(macro particle)/(turn)$\times$1000 (macro particles)
$\times$20(turns)$\times$injection efficiency(\%)$\times$90(\%)
$\times$80(\%)$\times$90(\%). The result is about 2.1$\times$10$^{10}$
 for the old tune case and 3.0$\times$10$^{10}$ for the new tune case
which are satisfying the required number of particles.

So far, space-charge effect is not included as mentioned before because
inclusion of the space charge requires more computing time. To see briefly the effect of the
space charge, 1000 macro particles are injected per each turn
representing $6\times 10^9$ protons.
Figure 7 shows the normalized horizontal phase space at the 1000$^{th}$ turn after the bump collapse
with the old tunes (1.73, 1.47) and old bump parameters 4.8 mrad, 20-turn injection,
and 20-turn linear pulse duration. It is found that the total injection efficiency decreases
to 26.9$\%$, approximately 20$\%$ reduction.

Figure 8 shows a similar plot with the new parameters. The injection efficiency in this case
becomes 39$\%$, again approximately 20$\%$ reduction from the case without space charge.

Figures 7 and 8 reveal structures of the phase space. This is believed to be higher-order resonances
caused by tune shifts in the presence of space charge effect~\cite{Hofmann} but it requires more study.

\section{CONCLUSIONS}
Increase in the efficiency for multi-turn injection of proton ions for a
cancer therapy synchrotron has been treated in this paper.
Transverse tunes are optimized by scanning tunes in the range $1 < \nu_x, \ \nu_y < 2$.
Then kick angle and number of turns for injection and the pulse duration of the bump magnet
were found which yielded most optimal injection efficiency.
It is found that injection efficiency can be enhanced by 17 percentage-point when space-charge effect is
not taken into account. In the presence of the space-charge effect, the efficiency is reduced by
20$\%$. Effect of magnetic errors such as multipole components, closed orbit distortion
is not treated in this paper and will be reported elsewhere.


\begin{acknowledgments}
This work is supported by Korea Heavy Ion Medical Accelerator (KHIMA)
project of Korea Institute of Radiological and Medical Sciences
(KIRAMS).
\end{acknowledgments}

\begin{figure}
\centering
\includegraphics[width=8cm, angle=90]{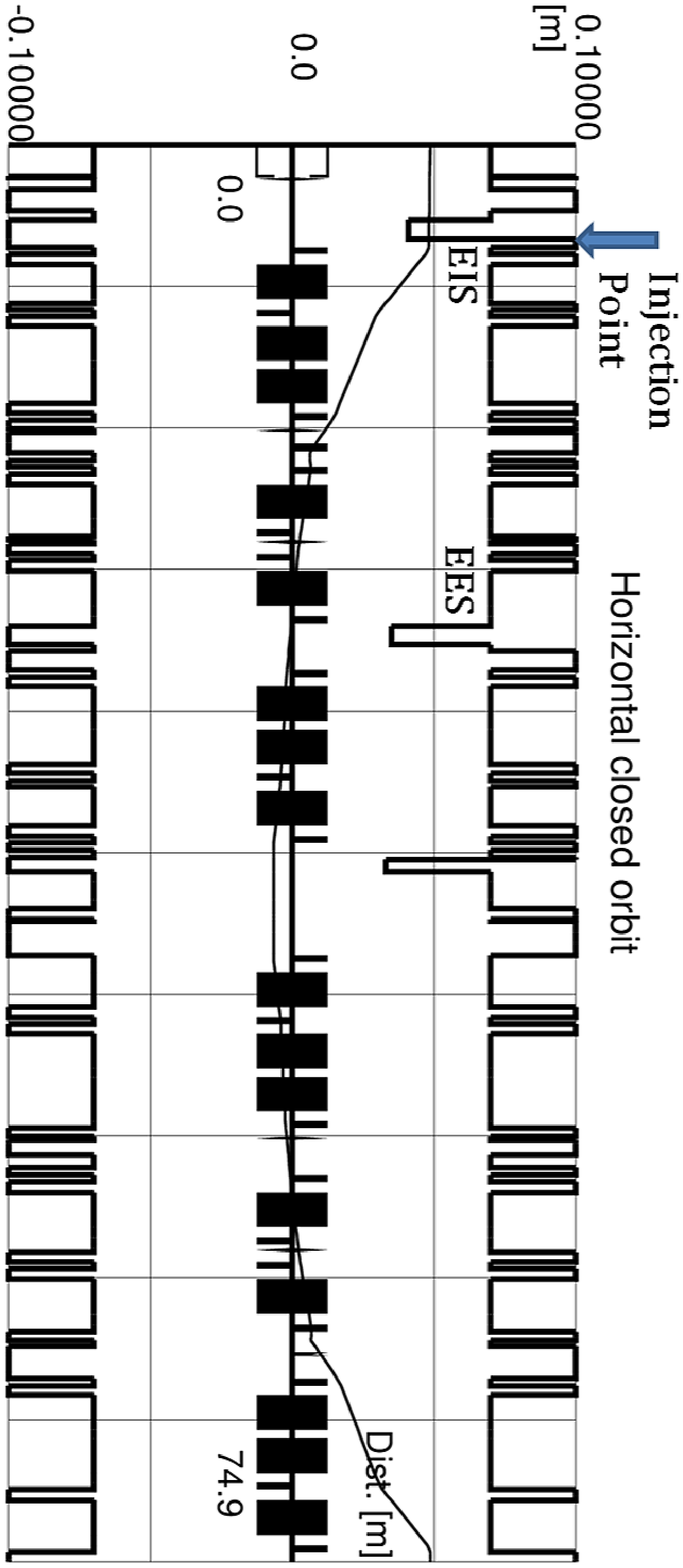}
\caption{[color on-line]The closed orbit path at the start of the injection } \label{fig.1}
\end{figure}

\begin{figure}
\centering
\subfloat[]{\includegraphics[width=8cm, angle=0]{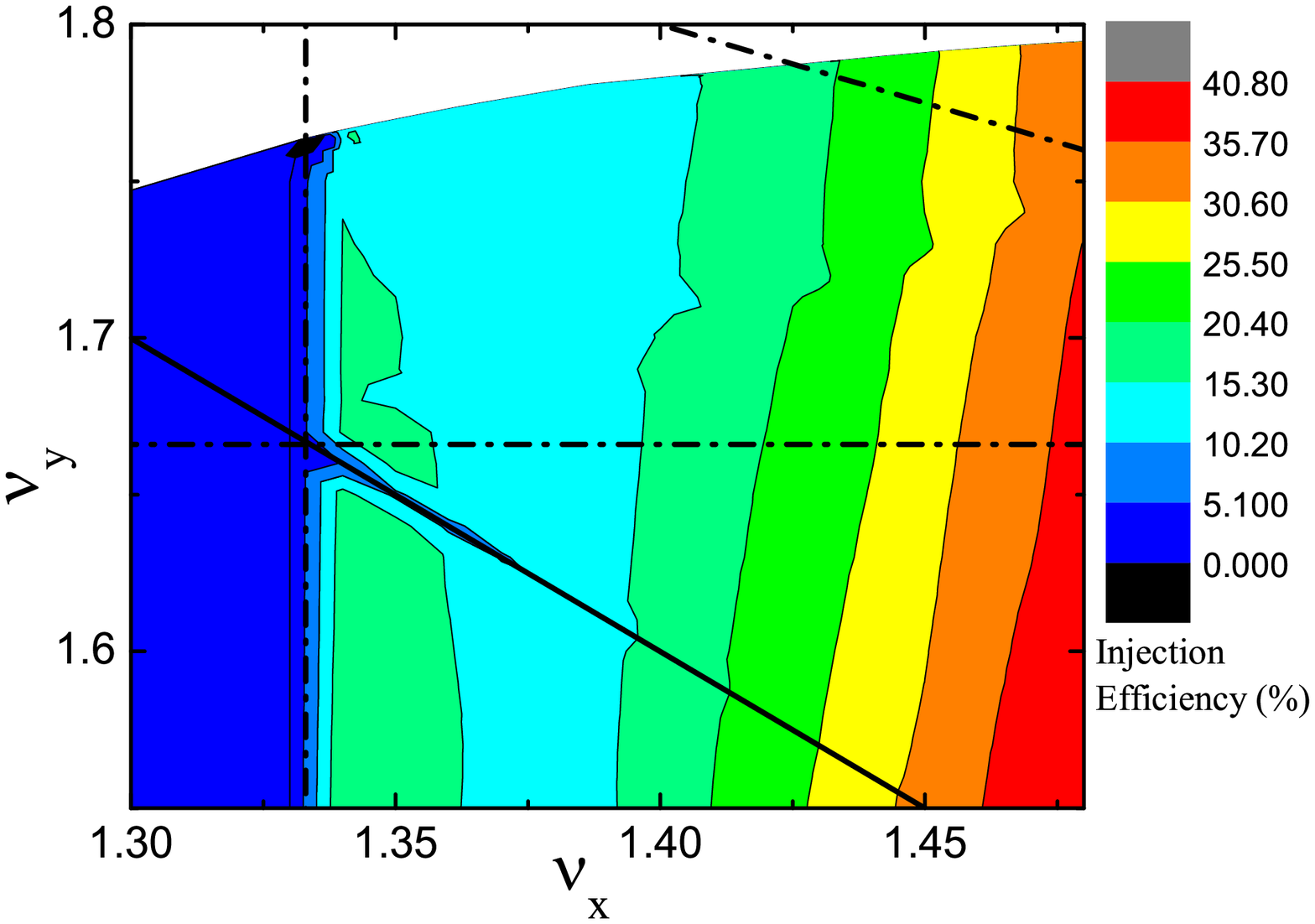}}
\subfloat[]{\includegraphics[width=8cm, angle=0]{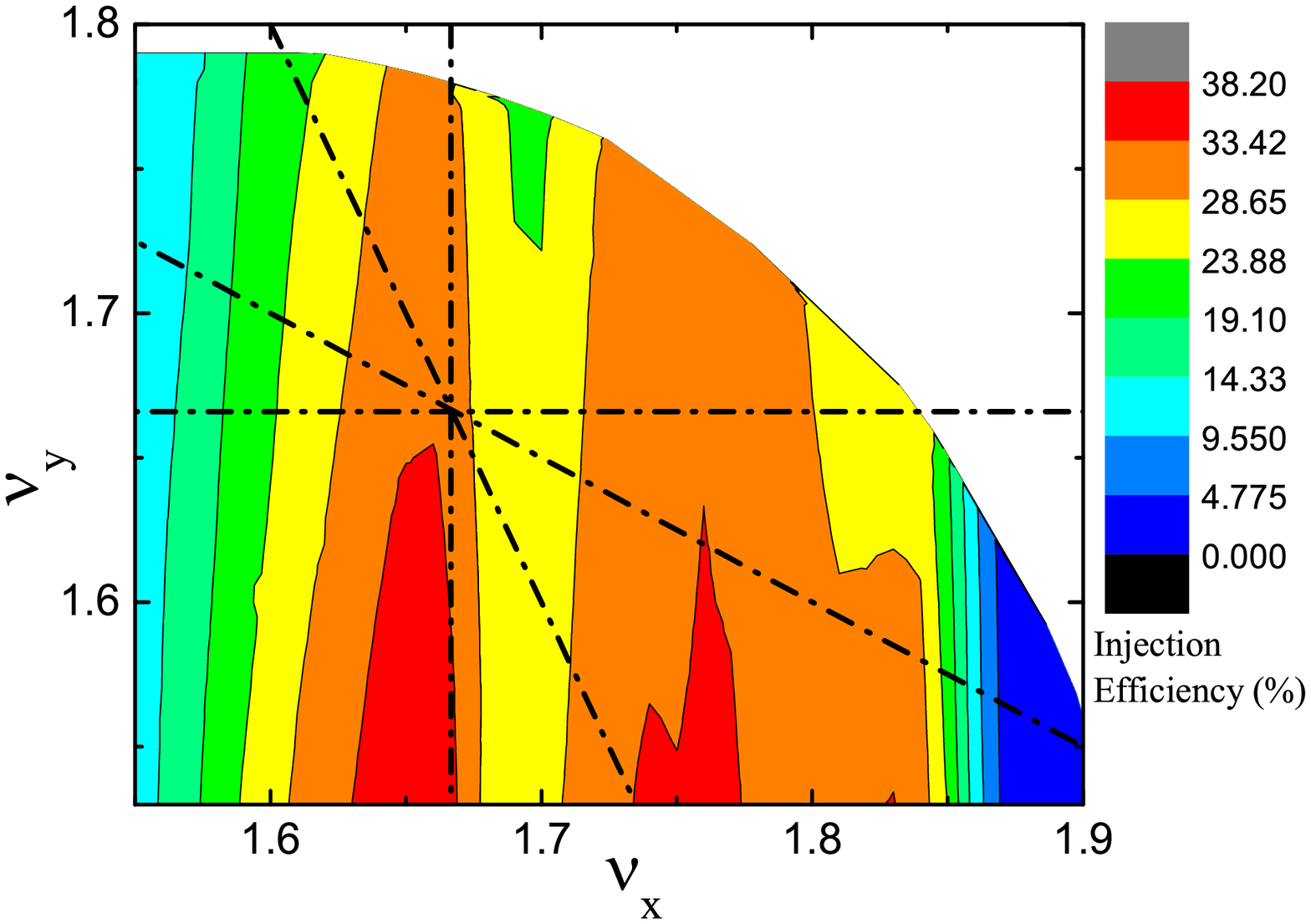}}\\
\subfloat[]{\includegraphics[width=8cm, angle=0]{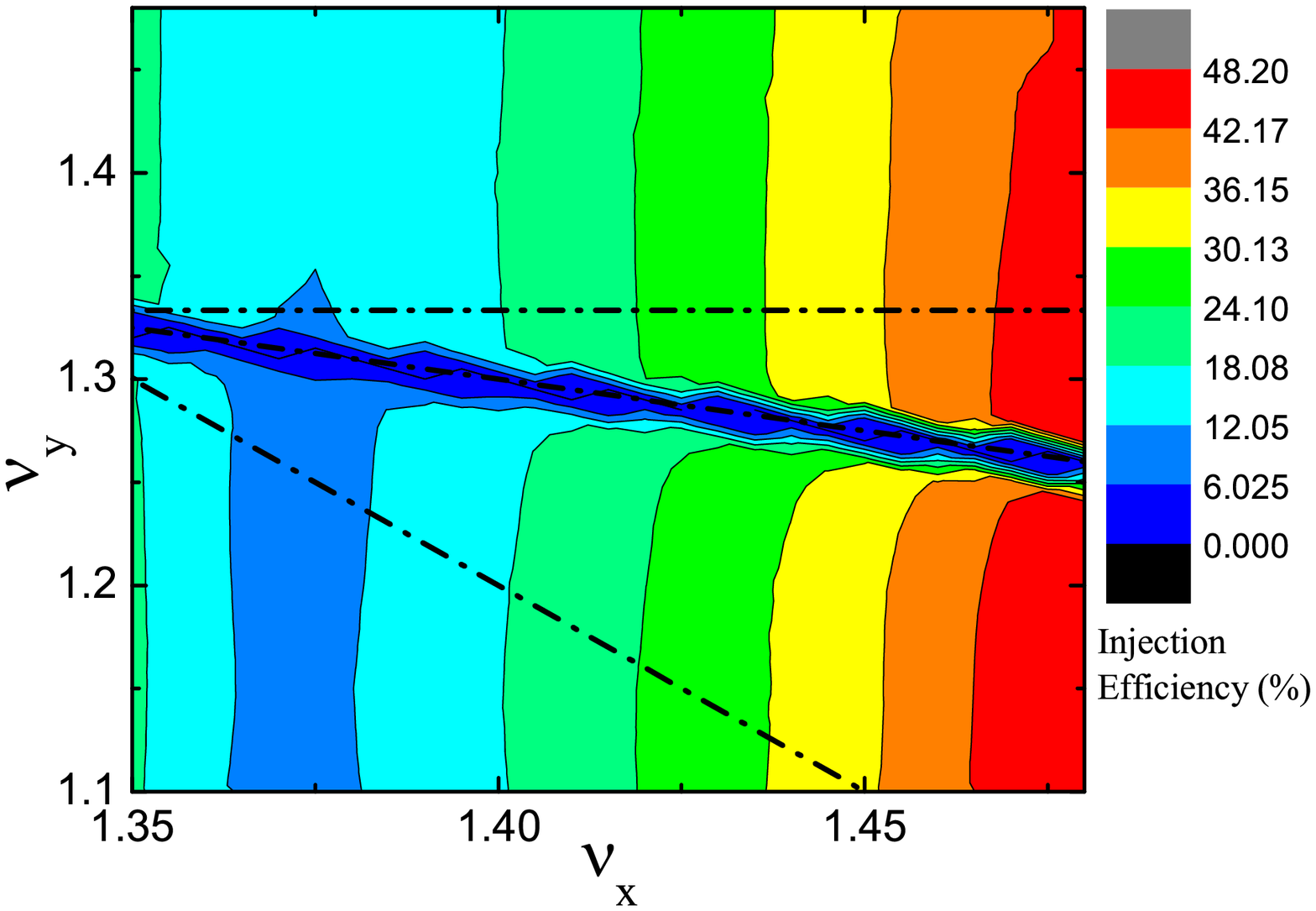}}
\subfloat[]{\includegraphics[width=8cm, angle=0]{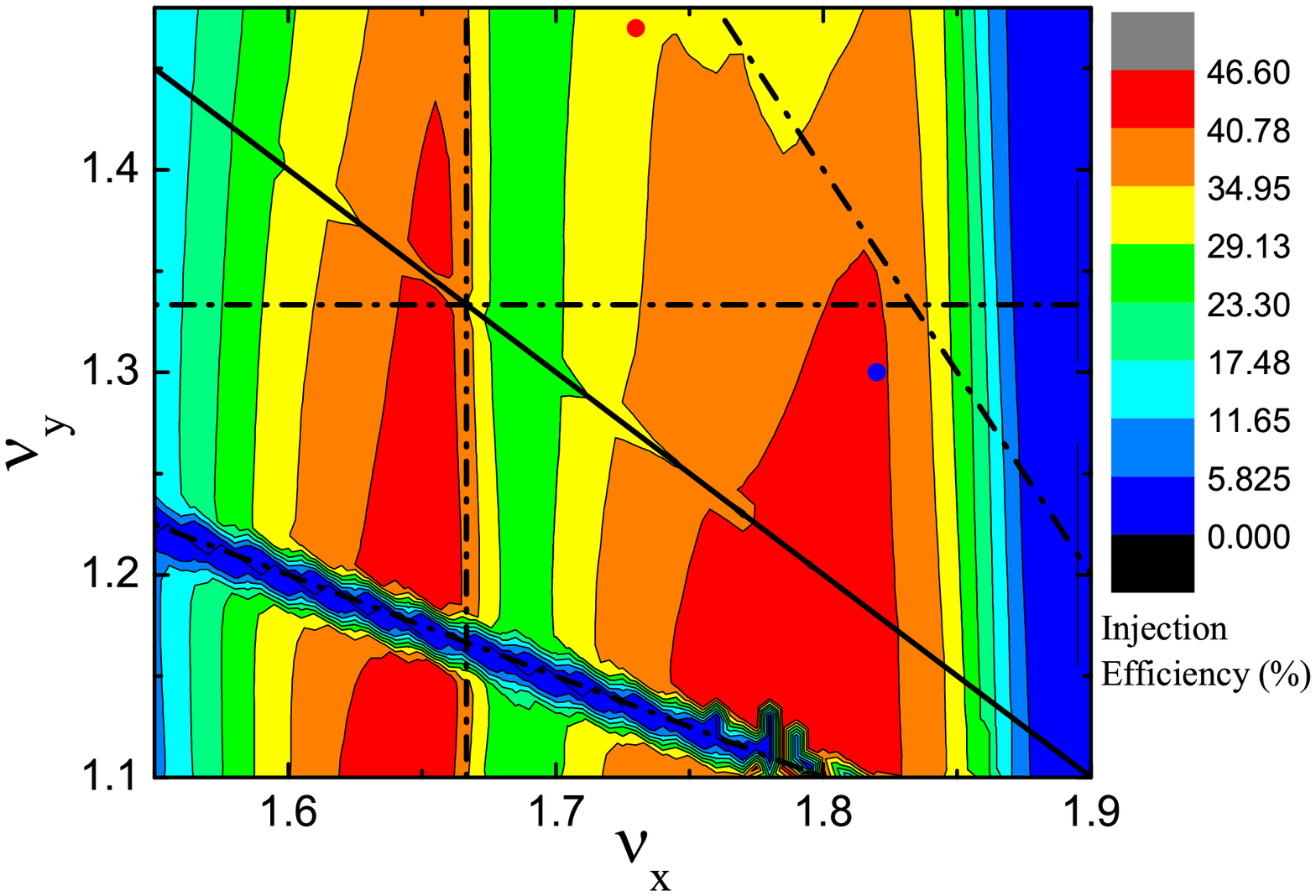}}
\caption{[color on-line]Foot print for injection efficiencies for $1 < \nu_{x} < 2$ and
$1 < \nu_{y} < 2$. Solid line show the second-order resonances and
dashed dot line indicate third-order resonances.} \label{fig.2}
\end{figure}

\begin{figure}
\centering
\subfloat[]{\includegraphics[width=8cm, angle=0]{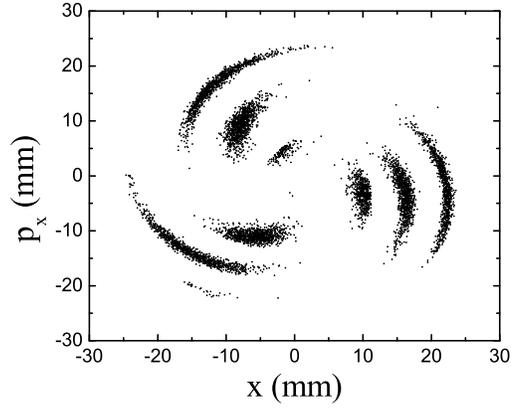}}\\
\subfloat[]{\includegraphics[width=8cm, angle=0]{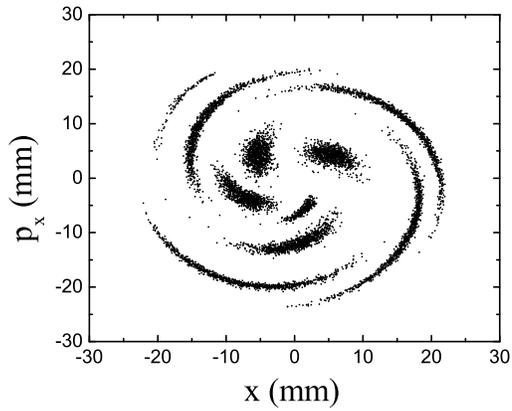}}
\caption{Normalized horizontal phase spaces at 1000 turn after
the bump collapse for (a) the initial tune (1.73, 1.47) and
(b) the new tune (1.83, 1.30). In both cases, the maximum kick angle is 4.8 mrad and
the bump pulse duration is 20 turns.} \label{fig.3}
\end{figure}

\begin{figure}
\centering
\includegraphics[width=8cm]{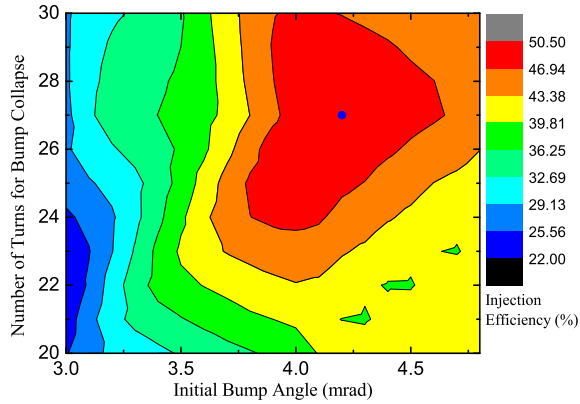}
\caption{[color on-line]Footprint for injection efficiencies  as functions of
initial kick angle of the bump magnet and the duration of the
linear pulse.
The dot indicates the selected value, 4.2 mrad and 27 turns} \label{fig.4}
\end{figure}

\begin{figure}
\centering
\includegraphics[width=8cm]{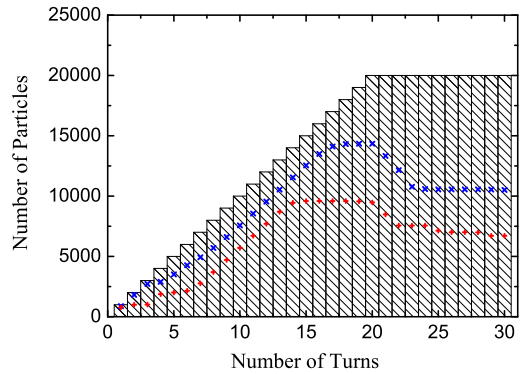}
\caption{[color on-line] Total number of accumulated particles as a function of
the number of turns
for old (plus marks) and new (cross marks) tunes and bump parameters.
The shaded bars indicate the number of desired accumulated macro particles
(i.e. number of particles for 100$\%$ injection efficiency) injected per each turn.
} \label{fig.6}
\end{figure}

\begin{figure}
\centering
\includegraphics[width=8cm]{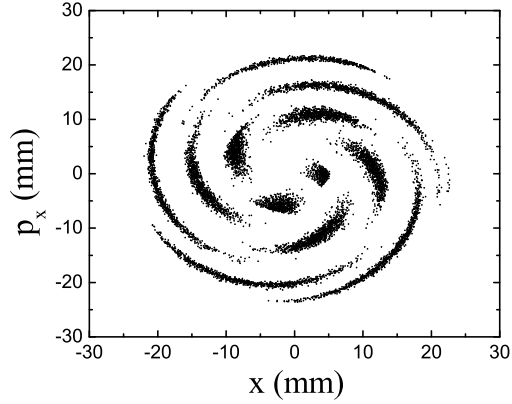}
\caption{Normalized horizontal phase space at 1000 turn after the bump collapse
with the new tunes (1.82, 1.30) and new bump parameters 4.2 mrad and 27 turns.
Space-charge effect is not included.} \label{fig.5}
\end{figure}

\begin{figure}
\centering
\includegraphics[width=8cm]{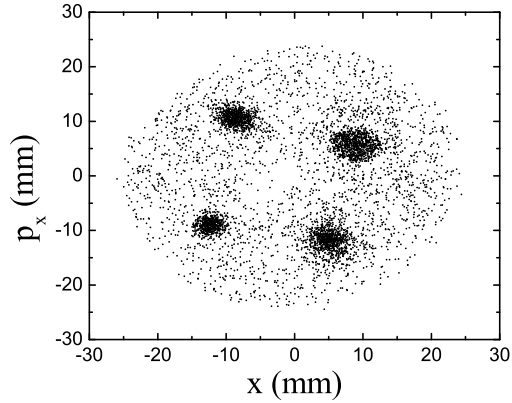}
\caption{Normalized horizontal phase space at 1000 turn after the bump collapse
with the old tunes (1.73, 1.47) and old bump parameters 4.8 mrad, 20-turn injection,
and 20-turn linear pulse duration.
Space-charge effect is included. One macro particle represents
$6\times 10^6$ protons and per each injection 1000 macro particles are injected.} \label{fig.7}
\end{figure}

\begin{figure}
\centering
\includegraphics[width=8cm]{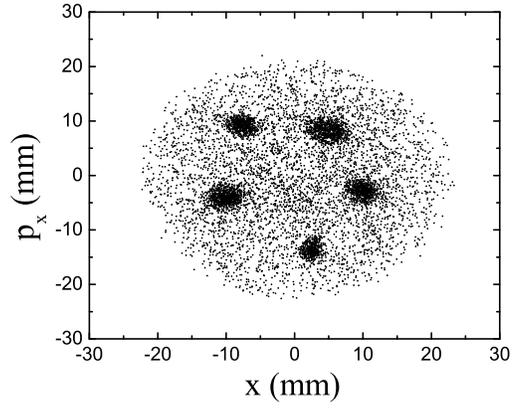}
\caption{Normalized horizontal phase space at 1000 turn after the bump collapse
with the new tunes (1.82, 1.30) and new bump parameters 4.2 mrad and 27 turns.
Space-charge effect is included. One macro particle represents
$6\times 10^6$ protons and per each injection 1000 macro particles are injected.} \label{fig.8}
\end{figure}

\end{document}